\documentclass[pra,twocolumn,superscriptaddress,10pt]{revtex4-1}
\usepackage{graphicx}
\usepackage{dcolumn}
\usepackage{color}
\usepackage{times}
\usepackage{bm}
\usepackage{amssymb}
\usepackage{amsmath}
\usepackage{epsfig}
\usepackage{epstopdf}
\usepackage{dsfont}
\usepackage{subfigure}
\usepackage{appendix}
\usepackage[colorlinks=true,citecolor=blue, urlcolor=blue, linkcolor=blue]{hyperref}
\begin{document}
\renewcommand{\thefootnote}{\fnsymbol {footnote}}
\title{Optimized entropic uncertainty relations for multiple measurements}

\author{Bo-Fu Xie}
\affiliation{School of Physics and Optoelectronic Engineering, Anhui University, Hefei
230601,  People's Republic of China}

\author{Fei Ming}
\affiliation{School of Physics and Optoelectronic Engineering, Anhui University, Hefei
230601,  People's Republic of China}

\author{Dong Wang} \email{dwang@ahu.edu.cn}
\affiliation{School of Physics and Optoelectronic Engineering, Anhui University, Hefei
230601, People's Republic of China}
\affiliation{CAS  Key  Laboratory  of Quantum  Information,  University  of  Science  and
Technology of China, Hefei 230026,  People's Republic of China}
\author{Liu Ye}  
\affiliation{School of Physics and Optoelectronic Engineering, Anhui University, Hefei
230601,  People's Republic of China}
\author{Jing-Ling Chen} \email{chenjl@nankai.edu.cn}
\affiliation{Theoretical Physics Division, Chern Institute of Mathematics, Nankai University, Tianjin 300071, People's Republic of China}

\date{\today}

\begin{abstract}{Recently, an entropic uncertainty relation for multiple measurements has been proposed by Liu {\it et al.} in [Phys. Rev. A 91, 042133 (2015)]. However, the lower bound of the relation  is not always tight with respect to different measurements. Herein, we improve the lower bound of the entropic uncertainty relation for multiple measurements, termed as simply constructed bound (SCB). We verify that the SCB is tighter than Liu {\it et al.}'s result
for arbitrary mutually unbiased basis measurements, which might play a fundamental and crucial role in practical quantum information processing.
Moreover, we  optimize the SCB by considering mutual information and the Holevo quantity, and
propose an optimized SCB (OSCB). Notably, the proposed bounds are extrapolations of the behavior of two measurements to a larger collection of measurements.
It is believed that our findings would shed light on entropy-based uncertainty relations in the multiple measurement scenario and will be beneficial for security analysis in quantum key distributions.
}
\end{abstract}

\maketitle

\section{Introduction}
There are some basic laws to characterize the quantum world  in compliance with quantum mechanics, of which the most well-known is the uncertainty
relation. The fundamental uncertainty relation is called the  Heisenberg uncertainty principle with regard to the position $\hat{x}$ and the momentum $\hat{p}$ in a given system $\rho$ \cite{Heisenberg}; it is expressed in the form of ${\Delta _\rho }\hat{x} {\Delta _\rho }\hat{p}\geq \hbar/2$, where the variance ${\Delta _\rho }\hat{x} = \sqrt {\left\langle {{\hat{x}^2}} \right\rangle_\rho   - {\left\langle \hat{x} \right\rangle }^2_\rho} $ and $\left\langle {\hat{x}} \right\rangle $ is  the expectation value of  $\hat x$. Subsequently, a more general formula
was derived by Kennard \cite{E. H. Kennard} and Robertson \cite{H. P. Robertson} for two arbitrary incompatible observables (say $\hat{R}$ and $\hat{S}$), whose lower bound is undesirably dependent on the systemic state. In the area of
information theory, entropy is considered a useful property to depict the state of a system. To eliminate Robertson's inequality weakness, Deutsch \cite{D. Deutsch} introduced Shannon entropy and proposed the entropic uncertainty relation, viz.
\begin{align}
H(\hat{R}) + H(\hat{S}) \ge 2\cdot{\log _2}\left(\frac{2}{{1 + \sqrt c }}\right).
\label{Eq.1}
\end{align}
$H(\hat{R})$ is the Shannon entropy given by $H(\hat{R}) =  - \sum\nolimits_i {{a_i}\log {a_i}}$, where $a_i = \langle\psi_i|\rho|\psi_i\rangle$ and $c = \mathop {\max }\limits_{ij} {\left| {\langle{\psi _i}|{\phi _j}\rangle } \right|^2}$ ($|\psi_i\rangle$, $|\phi_j\rangle$ are the eigenstates of $\hat{R}$ and $\hat{S}$, respectively). The lower bound is clearly not dependent on the state of the measured system. Subsequently, Kraus \cite{K. Kraus} as well as Maassen and Uffink \cite{H. Maassen} achieved a significant improvement on
\begin{align}
H(\hat{R}) + H(\hat{S}) \ge  - {\log _2}c=:q_{MU},
\label{Eq.2}
\end{align}
reflecting that the uncertainty will be the largest while measuring two arbitrary mutually unbiased observables.

Recently, a quantum-memory-assisted entropic uncertainty relation (QMA-EUR)  \cite{Joseph,M. Berta,Wang} has been presented, which has also been verified by several well-designed experiments \cite{C. F. Li,R. Prevedel}. To illustrate this new relation, we can consider a so-called quantum uncertainty game with two players Alice and Bob. Initially, Bob produces two particles $A$ and $B$ correlated in a quantum state and sends $A$ to Alice. Then, Alice randomly chooses $\hat{R}$ or $\hat{S}$ to measure on his particle and  tells Bob his measurement choice.  Bob can guess Alice's conclusion with the minimal uncertainty bounded by the proposed QMA-EUR.  Theoretically, the uncertainty relation in the presence of quantum memory can be expressed as \cite{Joseph,M. Berta}
\begin{align}
S(\hat{R}\left| B \right.) + S(\hat{S}\left| B \right.) \ge  - {\log _2}c + S(A|B),
\label{Eq.3}
\end{align}
where $S(\hat{R}\left| B \right.) = S({\rho ^{\hat{R}B}}) - S({\rho ^B})$  denotes the conditional von Neumann entropy of the post-measurement states. From the above relation in Eq. (\ref{Eq.3}), we note several interesting points: (i) one can  predict accurately the measurement's outcomes when $A$ and $B$ are maximally entangled, because of $S(A|B)=\log _2c=-\log_2 d$ with $d$ being the systemic dimension. {(ii) If particles $A$ and $B$ are disentangled, they can still be classically correlated as in a separable mixed state, and Eq. (\ref{Eq.3}) will still be a useful improvement over Eq. (\ref{Eq.2}). This corresponds to the case of a quantum uncertainty relation in the presence of a classical memory.}
 (iii) In the absence of the quantum memory, we obtain $H(\hat{R}) + H(\hat{S}) \ge  q_{MU}+S(\rho)$, whose bound is  stronger compared with Maassen and Uffink's bound in Eq. (\ref{Eq.2}) due to $S(\rho)\geq0$.

Afterwards, some authors proposed several tighter lower bounds of QMA-EUR by imposing different quantities \cite{A. K. Pati,F. Adabi,P. J. Coles, Fei}. Pati  \emph{et al.} \cite{A. K. Pati} verified that quantum correlation and classical correlation can strengthen the lower bound of QMA-EUR. Besides, Coles and Piani \cite{P. J. Coles} optimized the uncertainty relation with quantum memory by considering the second largest overlap of the two arbitrary observables to be measured. Recently, Adabi \emph{et al.} \cite{F. Adabi} optimized the lower bound of QMA-EUR by using the mutual information, which is expressed as
\begin{align}
S(\hat{R}\left| B \right.) + S(\hat{S}\left| B \right.) \ge  - {\log _2}c + S(A\left| {B)} \right. + \max \{0,\delta \},
\label{Eq.4}
\end{align}
where $\delta  = {\cal{I}}(A:B) - \left[ {\cal{I}}(\hat R:B) + {\cal{I}}(\hat S:B) \right]$, { ${\cal{I}}(A:B) = S\left( {{\rho _A}} \right) + S\left( {{\rho _B}} \right) - S\left( {{\rho _{AB}}} \right)$ stands for the mutual information, ${\cal{I}}(\hat R:B) = S \left({{\rho _{\hat{R}}}}\right)+ S\left( {{\rho _B}} \right) - S\left( {{\rho _{\hat{R}B}}} \right)$ and ${\cal{I}}(\hat S:B) = S\left( {{\rho _{\hat{S}}}} \right) + S\left( {{\rho _B}} \right) - S\left( {{\rho _{\hat{S}B}}} \right)$ are defined as the Holevo quantity.}
In particular, QMA-EUR has received wide and extensive attention owing to its powerful applications in implementing quantum tasks, such as quantum metrology \cite{C. S. Yu}, entanglement witness \cite{M. Berta}, quantum key distribution \cite{N. J. Cerf, Fei}, quantum phase transition \cite{X. M. Liu}, quantum cryptography \cite{F. Dupuis, F. Grosshans}, and EPR steering \cite{J. Schneeloch}. From an experimental perspective, significant progress on QMA-EUR has been achieved by some groups \cite{W. C. Ma, Z. X. Chen, W. M. Lv, H. Y. Wang, ZX.Chen,ding}.

Moreover, pursuing the EUR versions for multiple measurements has become very interesting after several excellent simulations and calculations based on two-measurement were reported  \cite{L. M. Hu,J. Feng,Z. Y. Xu.,L. M. Hu.,Z. Y. Xu,  Y. Lim,  M. N. Chen, P. J. Coles11,Dong5}. Recently,  Liu {\it et al.}  proposed an entropic uncertainty relation for multiple measurements as \cite{S. Liu}
\begin{align}
\sum\limits_{m=1}^N {H({\Hat{M}_m}\left| {B)} \right. \ge  - {{\log }_2}(f) + (N - 1)S(A\left| {B)} \right.} :=U_{\rm LMF},
\label{Eq.5}
\end{align}
where $f\!\! =\!\! \mathop {\max }\limits_{{i_N}}\!\! \left\{\! {\sum\limits_{{i_2} \!\sim {i_{N\! - \!1}}}\!\!\! {\mathop {\max }\limits_{{i_1}}\! \left[\! {c\!\left(\! {u_{{i_1}}^1\!,u_{{i_2}}^2}\! \right)}\! \right]\prod _{m\! = \!2}^{N\! - \!1}\left[\! {c\left(\! {u_{{i_m}}^m\!,u_{{i_{m\! +\! 1}}}^{m\! +\! 1}}\! \right)}\! \right]} }\! \right\}$ and ${u_{{i_m}}^m}$ is the $i$-th
eigenvector of the operator $\Hat{M}_m$ \cite{Y. L. Xiao}. However, such a lower bound is not tight for all incompatible measurements. The aim of this study is to explore the improved QMA-EUR with multiple measurements. Specifically, we present a simply constructed bound (SCB) of QMA-EUR for multiple measurements. Furthermore, we optimize the SCB by taking into account the systemic mutual information. It is interesting that our proposed bounds are stronger than Liu {\it et al.}'s bound in Eq. (\ref{Eq.5}) with regard to arbitrary mutually unbiased bases (MUB) measurements.

\section{Two lower bounds for multiple measurements}\vskip 0.2cm


 Similar to the additivity of linear uncertainty relations for local measurements \cite{S. Wehner, R. Schwonnek}, we extend the entropic
uncertainty relation in Eq. (\ref{Eq.3}) to an improved uncertainty relation for the case of multiple measurements  as
\begin{align}
\sum\limits_{m=1}^N {S({\Hat{M}_m}\left| B \right.) \ge  - \frac{1}{{N - 1}}{{\log }_2}\left( {\prod\limits_{i=1}^{C_N^2} {{c_i}} } \right)}  &+ \frac{N}{2}S(A | B )\nonumber\\
&:=U_{\rm SCB},
\label{Eq.6}
\end{align}
where ${C_N^2}=N(N-1)/2$ is the combinatorial number of choosing two different measuring operators from above $N$ different operators. $c_i=\mathop {\max }\limits_{j,k}|\langle a_j^m|a_k^n \rangle|^{2}$, where $|a_j^m\rangle$, $|a_k^n\rangle$ represent the eigenvectors of $\Hat{M}_m$ and $M_n$. $U_{SCB}$ denotes  the right-hand side of Eq. (\ref{Eq.6}), called the simply constructed bound.  \vskip 0.2cm
{\bf{Theorem 1.}} The SCB presented in  Eq. (\ref{Eq.6}) is tighter than Liu {\it et al.}'s bound in Eq. (\ref{Eq.5}) with respect to arbitrary multiple mutually unbiased bases measurements.  \vskip 0.2cm
{\bf{Proof.}} First, let us review the characteristics of mutually unbiased bases (MUB). We assume there are two groups of orthonormal bases $\{|P_{i}\rangle\}_{i=1, 2,\cdots, d}$ and $\{|Q_{j}\rangle\}_{j=1, 2,\cdots, d}$ in a complex Hilbert space of dimension $d$. Then, both of them are called mutually unbiased bases (MUB) if
\begin{align}
|\langle P_i|Q_j\rangle|=\frac{1}{\sqrt{d}}
\label{Eq.7}
\end{align}
holds for all $i$ and $j$ \cite{S. Designolle}. Here, we can derive Liu {\it et al.}'s bound for MUB measurements in Eq. (\ref{Eq.5}). Hereafter, we take $N$ groups of $d$-dimensional mutually unbiased bases for measurement, viz.  $c\left(\! {u_{{i_m}}^m\!,u_{{i_{m\! +\! 1}}}^{m\! +\! 1}}\! \right)=\frac{1}{d}$, where the form of $f$ will collapse into $f=\frac{1}{d}\times d^{N-2}\times {\frac{1}{d^{N-2}}}=\frac{1}{d}$. Then, the right bound of Eq. (\ref{Eq.5}) becomes
\begin{align}
\log_2d+(N-1)S(A|B):=RB1.
\label{Eq.8}
\end{align}

Our SCB's general form is shown in Eq. (\ref{Eq.6}). Similarly, if we consider the MUB measurement, viz. $c_i=\mathop {\max }\limits_{j,k}|\langle a_j^m|a_k ^n\rangle|^{2}=\frac{1}{d}$, the right bound of SCB can be expressed as
\begin{align}
\frac{N}{2}\log_2d+\frac{N}{2}S(A|B):=RB2.
\label{Eq.9}
\end{align}
To compare these two  bounds, we can calculate the difference between them as
\begin{align}
RB2-RB1=(\frac{N}{2}-1)\left(\log_2d-S(A|B)\right).
\label{Eq.10}
\end{align}
With regard to $S(A|B)$, we have
\begin{align}
S(A|B)&=S(A, B)-S(B)\nonumber\\
&\leq S(A)+S(B)-S(B)=S(A)\nonumber\\
&\leq \log_2d.
\label{Eq.11}
\end{align}
Thus, due to $N\geq 3$ and combining Eqs. (\ref{Eq.10}) and (\ref{Eq.11}),  we determine  that $RB2\geq RB1$ is always satisfied. Additionally,  Eq. (\ref{Eq.10}) indicates that our bound will recover  Liu {\it et al.}'s result as shown in Eq. (\ref{Eq.5}) for $N=2$. As a result,
we can conclude that our SCB is tighter than Liu {\it et al.}'s bound for any multiple MUB measurements. \vskip 0.2cm

${\bf{Theorem\ 2.}}$ By taking into account the mutual information {and the Holevo
quantity} in the SCB, one can attain a stronger bound of entropic uncertainty relations in the regime of multiple measurements, which is given by
\begin{align} \begin{split}
\sum\limits_{m=1}^N {S({\Hat{M}_m}| B) \geq  - \frac{1}{{N - 1}}{{\log }_2}\left( {\prod\limits_{i=1}^{C_N^2} {{c_i}} } \right)}  +
\frac{N}{2}S(A| B) \\
 + \max\{0,\delta_m\}:=U_{\rm OSCB},
\label{Eq.12}
\end{split}\end{align}
where $\delta_m=\frac{N}{2}{\cal {I}}(A:B) - \sum\limits_{m=1}^N {{\cal{I}}({\Hat{M}_m}: B)}$, {${\cal {I}}(A:B)$ and ${\cal{I}}({\Hat{M}_m}: B)$ denote the the mutual information and the Holevo
quantity, respectively}. ${C_N^2}$ and $c_i$ are the same as the definitions in Eq. (\ref{Eq.6}),
and $U_{\rm OSCB}$ denotes the optimized SCB (OSCB).  \vskip 0.2cm

${\bf{Proof.}}$ Assume that there are two arbitrary measurement operators $\hat{M}_i$ and $\Hat{M}_j$ ($i\neq j$), and then, the preparation uncertainty can be expressed by
\begin{align}
&S(\hat{M}_i|B)+S(\Hat{M}_j|B)\nonumber\\
&=S(\hat{M}_i)-{\cal {I}}(\hat{M}_i:B)+S(\Hat{M}_j)-{\cal {I}}(\Hat{M}_j:B)\nonumber\\
&\geq \log_2\frac{1}{c}+S(A)-[{\cal{I}}(\hat{M}_i:B)+{\cal{I}}(\Hat{M}_j:B)]\nonumber\\
&=\!\log_2\frac{1}{c}\!+\!S(A|B)\!+\!{\cal{I}}(A:B)\!-\![{\cal{I}}(\hat{M}_i:B)\!+\!{\cal{I}}(\Hat{M}_j:B)],
\label{Eq.13}
\end{align}
considering that $S(\hat{M}_i)+S(\Hat{M}_j)\geq \log_2\frac{1}{c}+S(A)$ and $S(A)=S(A|B)+{\cal{I}}(A:B)$. For $i, j\in \{1, 2, 3, \cdots, N\}$ ($i\neq j$), we can obtain  $C_N^2$ analogous relations as derived in Eq. (\ref{Eq.13}). Therefore, by adding these $C_N^2$ inequalities, we can attain a new expression as
\begin{align}
(N\!-\!1)\!\sum\limits_{m=1}^N &S(\Hat{M}_m|B)\!\geq\!-{{\log }_2}\left( {\prod\limits_{i=1}^{C_N^2} {{c_i}} } \right)\!\!+\!\!\frac{N(N\!-\!1)}{2}\!S(A|B)\nonumber\\
&+\!\!\frac{N(N-1)}{2}{\cal{I}}(A:B)\!-\!(N\!-\!1)\sum\limits_{m=1}^N {\cal{I}}(\Hat{M}_m\!:\!B).
\label{Eq.14}
\end{align}
By simplifying the above formula, we  have
\begin{align}
\sum\limits_{m=1}^N {S({\Hat{M}_m}| B) \geq  - \frac{1}{{N - 1}}{{\log }_2}\left( {\prod\limits_{i=1}^{C_N^2} {{c_i}} } \right)}  +
\frac{N}{2}S(A| B)\nonumber \\
 + \max\{0,\delta_m\}:=U_{\rm OSCB},
\label{Eq.15}
\end{align}
which recovers Eq. (\ref{Eq.12}) in Theorem 2, where $\delta_m=\frac{N}{2}{\cal {I}}(A:B) - \sum\limits_{m=1}^N {{\cal{I}}({\Hat{M}_m}: B)}$. Based on the inequality, we derive the following conclusions: (i) the OSCB in Eq. (\ref{Eq.12})  must be tighter than the SCB in Eq. (\ref{Eq.6}) when $\delta_m>0$; (ii) if $\delta_m\leq0$, the OSCB will recover the SCB. With this consideration, we can declare that our proposed OSCB is always stronger compared with the SCB. Importantly, we argue that our OSCB must be stronger than both Liu {\it et al.}'s bound in Eq. (\ref{Eq.5}) and the SCB  for arbitrary MUB measurements.

\section{Examples}
To verify our findings, we consider several typical examples based on different MUB measurements in systems with various pure and mixed states including ensembles' state and random states.
\subsection{Werner states}
\begin{figure}
\centering
\includegraphics[width=7.7cm]{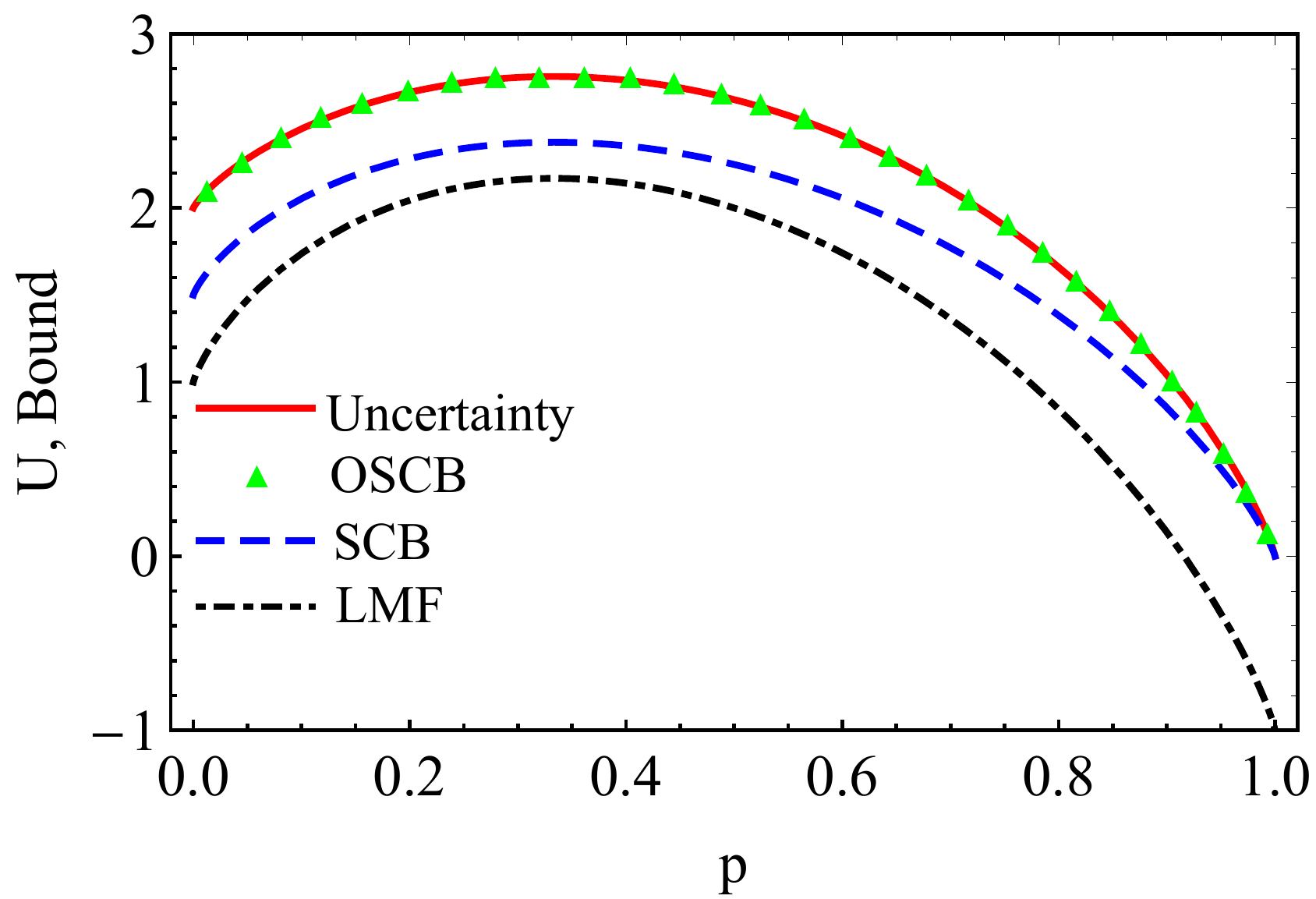}
\caption{Uncertainty and the lower bounds versus { Werner} state's purity $p$. The red solid line represents the entropic uncertainty (the left hand-side of Eq. (\ref{Eq.6})), the blue dashed line shows our SCB (the right-hand side of Eq. (\ref{Eq.6})) and {the black dot-dashed line} shows Liu {\it et al.}'s result (the right hand-side of Eq. (\ref{Eq.5})). {The green triangle} represents our OSCB (the right-hand of Eq. (\ref{Eq.12})).}
\label{figure1}
\end{figure}

First, we consider a two-qubit Werner state, which can be expressed as
\begin{align}
\rho_{AB}=\frac{1-p}{4}\mathds{1}_{A}\otimes \mathds{1}_{B}+p|\Psi^-\rangle_{AB}\langle\Psi^-|,
\label{Eq.16}
\end{align}
where $0\leq p\leq1$ denotes the purity of the state, and $|\Psi^-\rangle_{AB}=\frac{1}{\sqrt{2}}(|01\rangle-|10\rangle)$ is one of the Bell states. Besides, we take into account three two-dimensional Pauli operators $\hat{{\sigma}}_{x}$, $\hat{{\sigma}}_{y}$ and $\hat{{\sigma}}_{z}$ as the incompatible measurements, which typically belong to MUB measurements. As shown in Fig. \ref{figure1}, our SCB
is tighter than Liu {\it et al.}'s bound. Specifically, Liu {\it et al.}'s bound will be less than $0$ with increasing $p$, which is a trivial result because the value of the uncertainty is nonnegative. Therefore, we can say that Liu {\it et al.}'s lower bound is not a good candidate for measuring the entropic uncertainty in this case, while our SCB and OSCB are competent for this purpose. Moreover, our OSCB is always synchronized with the  uncertainty as displayed in Fig. \ref{figure1}, which reflects that our OSCB is optimal compared with the one in Ref. \cite{S. Liu}  and the SCB  presented in Eq. (\ref{Eq.6}).

\subsection{Bell diagonal states}
As another illustration, we consider a set of two-qubit states with maximally mixed marginal states, which is represented as
\begin{align}
\rho_{AB}=\frac{1}{4}\left(\mathds{1}\otimes \mathds{1}+\sum\limits_{i,j}^3 {{\omega _{i,j}}{\hat{\sigma} _i} \otimes {\hat{\sigma} _j}}\right),
\label{Eq.17}
\end{align}
where $\hat{\sigma}_{i}(i=1, 2, 3)$ are the Pauli matrices. According to the singular-value-decomposition theorem, the above state of two Dirac field modes transform to the following form of the Bell-diagonal  state:
\begin{align}
\rho_{AB}=\frac{1}{4}\left(\mathds{1}\otimes \mathds{1}+\sum\limits_{i}^3{r_{i}\hat{\sigma}_{i}\otimes\hat{\sigma}_{i}}\right).
\label{Eq.18}
\end{align}
This density matrix is positive if $\vec{r}=(r_{1}, r_{2}, r_{3})$ belongs to a tetrahedron defined by the set of vertices $(-1, -1, -1), (-1, 1, 1), (1, -1, 1)$ and $(1, 1, -1)$. Here, we
set $r_{1}=1-2p, r_{2}=r_{3}=-p$; thus, the state in Eq. (\ref{Eq.18}) becomes
\begin{align}
\rho_{AB}=p|\psi^-\rangle\langle\psi^-|+\frac{1-2p}{2}(|\psi^+\rangle\langle\psi^+|+|\phi^+\rangle\langle\phi^+|),
\label{Eq.19}
\end{align}
where $|\phi^\pm\rangle=\frac{1}{\sqrt{2}}(|00\rangle\pm|11\rangle), |\psi^\pm\rangle=\frac{1}{\sqrt{2}}(|01\rangle\pm|10\rangle)$ are the Bell states and $p$ represents the purity {of the state}.

\begin{figure}
\centering
\includegraphics[width=7.7cm]{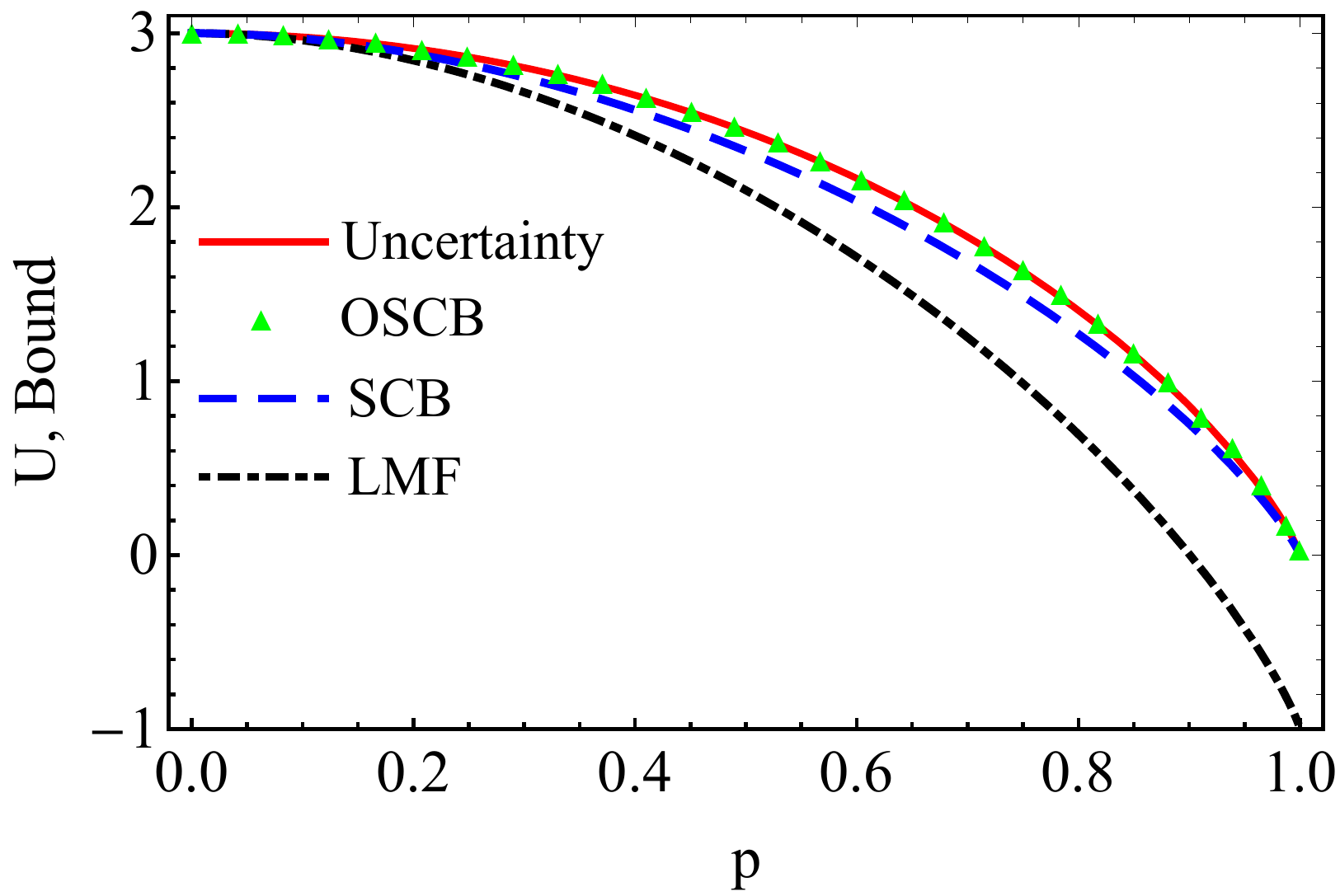}
\caption{Uncertainty and the lower bounds versus the Bell-diagonal state's purity $p$. The red solid line represents the entropic uncertainty (the left hand-side of Eq. (\ref{Eq.6})), the blue dashed line shows our SCB (the right-hand side of Eq. (\ref{Eq.6})) and {the black dot-dashed  line} shows Liu {\it et al.}'s result (the right hand-side of Eq. (\ref{Eq.5})). {The green triangle}  represents our OSCB (the right-hand of Eq. (\ref{Eq.12})).}
\label{figure2}
\end{figure}

Fig. \ref{figure2} plots the uncertainty, our lower bounds (SCB and OSCB) and  Liu {\it et al.}'s bound (LMF) as a function of the state's purity $p$ for three MUB measurements $\hat{\sigma}_{x}, \hat{\sigma}_{y}$ and $\hat{\sigma}_{z}$. It shows that our SCB is tighter than Liu {\it et al.}'s bound for arbitrary $p$, and Liu {\it et al.}'s bound is less than 0 for relatively large $p$, leading to an undesirable result. Moreover, as an optimized bound, our OSCB always coincides with the entropic uncertainty and is stronger compared with SCB and  Liu {\it et al.}'s bound.

\subsection{Random two-qubit states}

\begin{figure*}
\begin{minipage}{8.5cm}
\centering
\includegraphics[width=7.5cm]{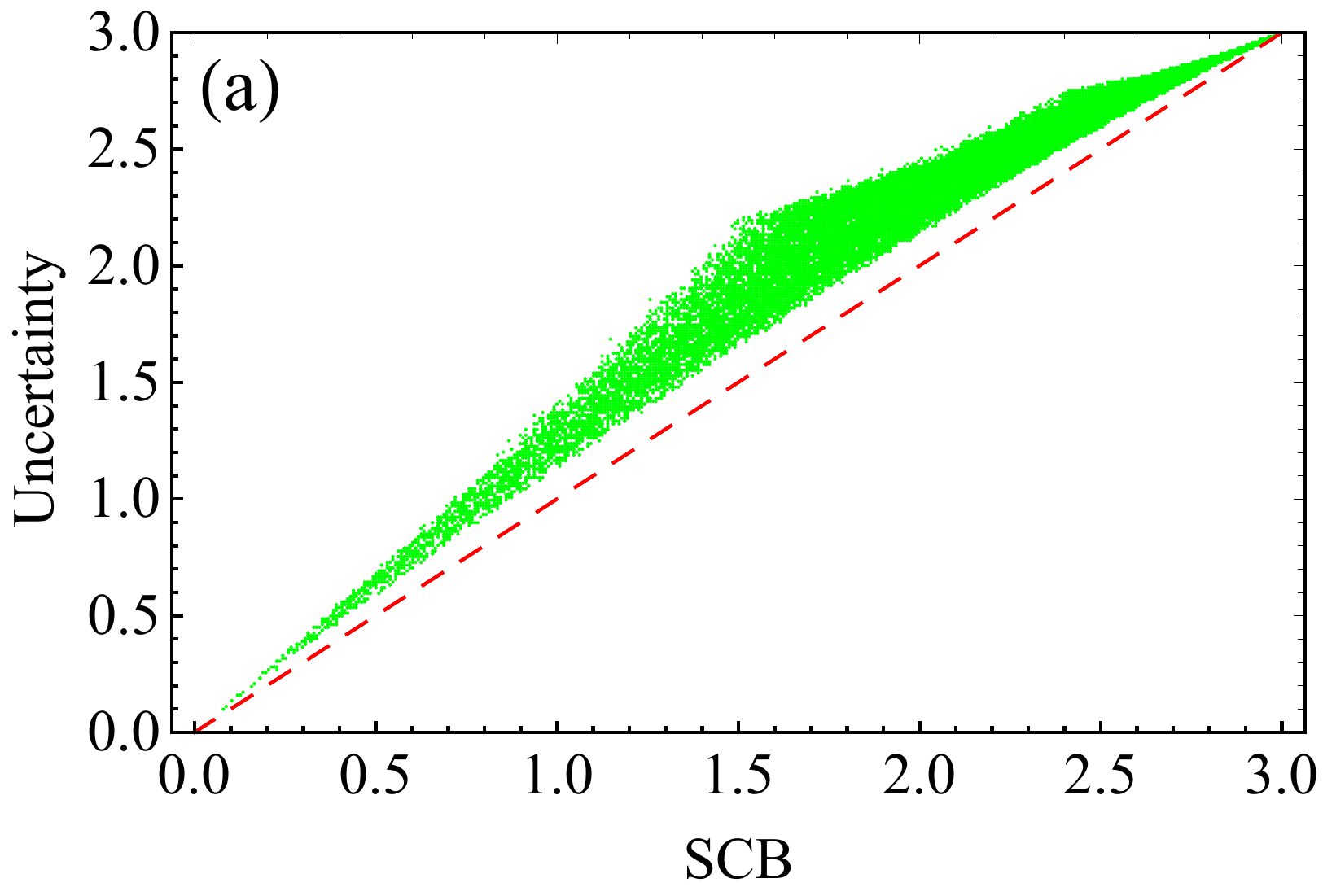}
\end{minipage}
\begin{minipage}{8.5cm}
\centering
\includegraphics[width=7.5cm]{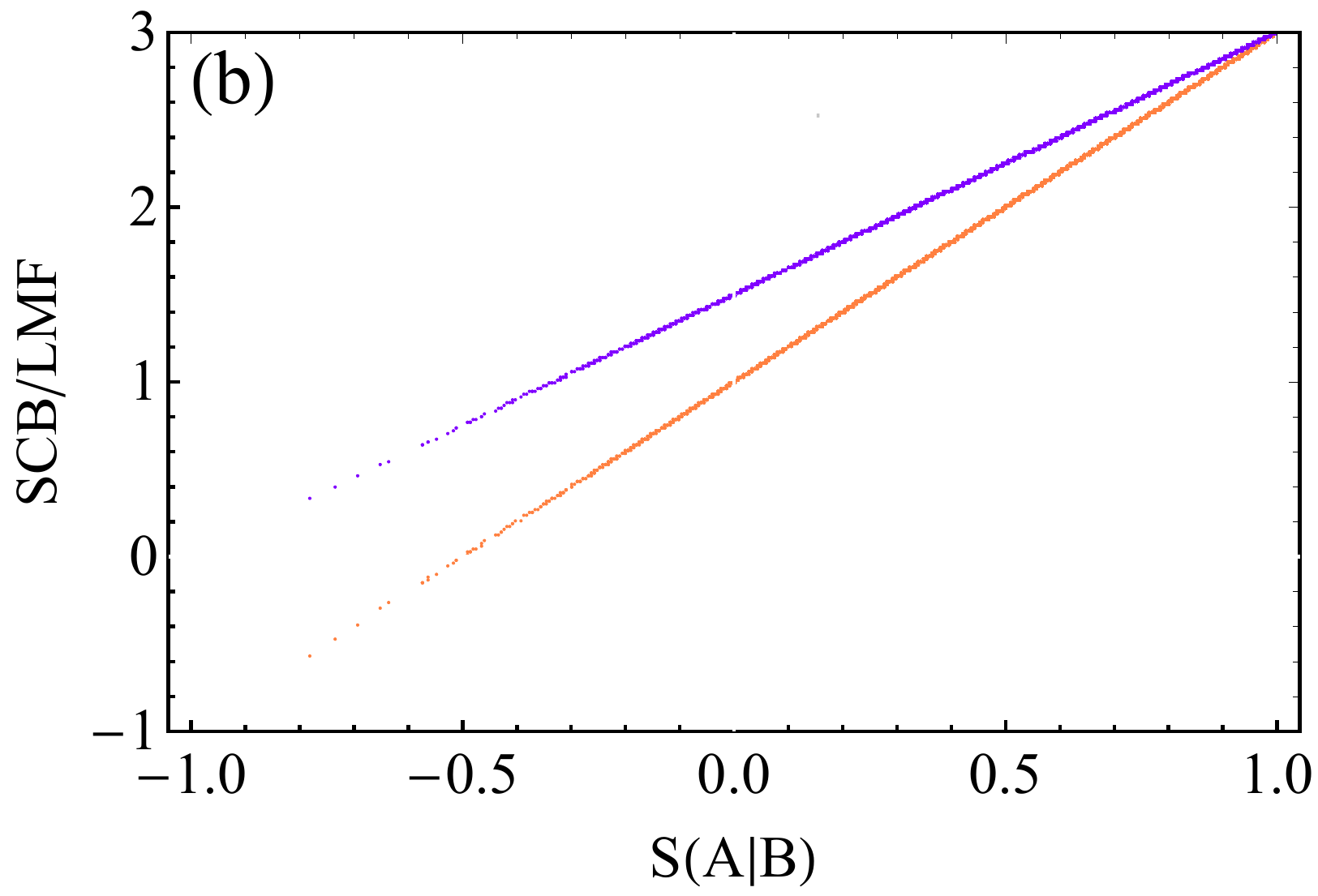}
\end{minipage}
\begin{minipage}{8.5cm}
\centering
\includegraphics[width=7.5cm]{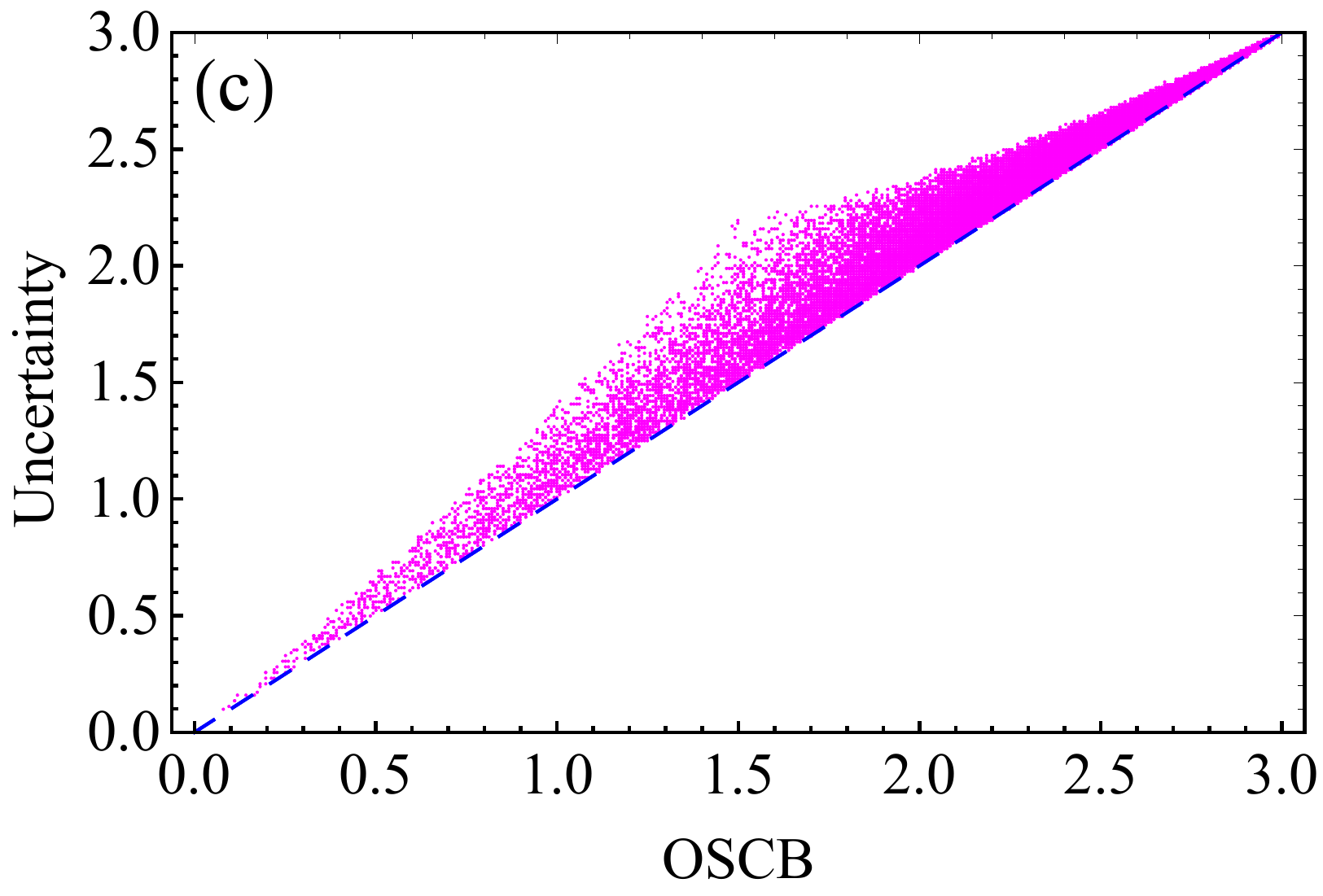}
\end{minipage}
\begin{minipage}{8.5cm}
\centering
\includegraphics[width=7.5cm]{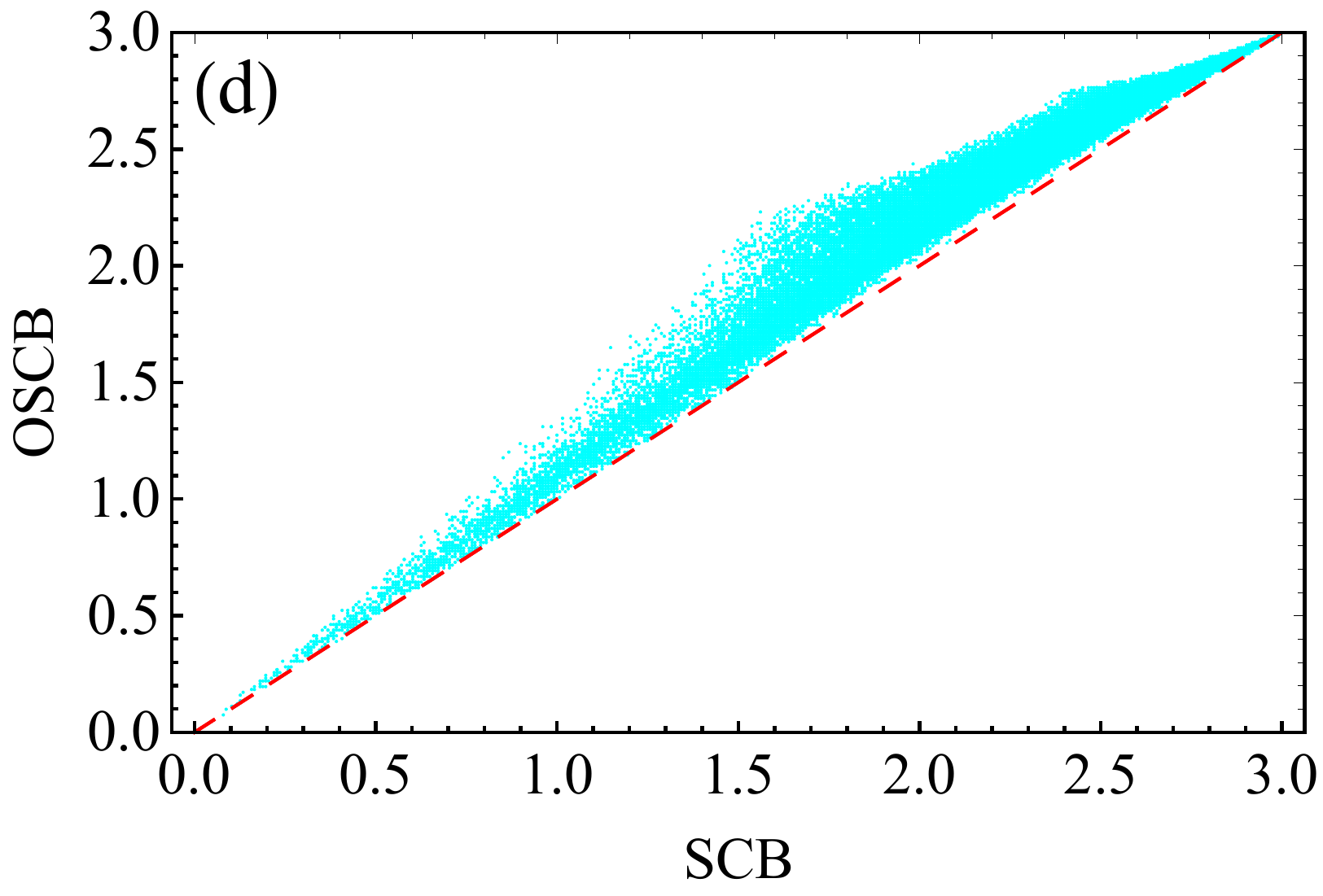}
\end{minipage}
\caption{{{The entropic uncertainty and three lower bounds under MUB multiple measurements (three two-dimensional Pauli operators $\hat{\sigma}_{x}, \hat{\sigma}_{y}, \hat{\sigma}_{z}$) for $10^6$ randomly generated two-qubit states. Graph (a): $x$ axis represents our simply constructed bound (SCB) in Eq. (\ref{Eq.6}) and $y$ axis denotes the entropic uncertainty in Eq. (\ref{Eq.6}). The red dashed line is the proportional function with a slope of unity, showing that the uncertainty is equal to SCB.  Graph (b): the orange points   stand for Liu {\it et al.}'s bound (LMF) in Eq. (\ref{Eq.5}), the purple points denote our proposed bound (SCB) in Eq. (\ref{Eq.6}), and $S(A|B)$ is the conditional von Neumann entropy of state $\rho_{AB}$.  Graph (c): $x$ axis  represents our optimized simply constructed bound (OSCB) in Eq. (\ref{Eq.12}), $y$ axis  represents the entropic uncertainty in Eq. (\ref{Eq.6}), and the blue dashed line is the proportional function with a slope of unity, showing that the uncertainty is equal to OSCB. Graph (d):  SCB represents the lower bound  in Eq. (\ref{Eq.6}) and OSCB denotes the lower bound in Eq. (\ref{Eq.12}). The red dashed line is the proportional function with a slope of unity, showing that OSCB is equal to SCB in this case.}}}
\label{figure3}
\end{figure*}

In order to better demonstrate our bounds' optimization, we consider random two-qubit states. First, let us introduce how to construct random two-qubit states. An arbitrary two-qubit state $\rho_{AB}$ can be decomposed by the state's eigenvalues $p_{n}$ and normalized eigenvectors $|\varphi_{n}\rangle$, viz. $\rho_{AB}=\sum\limits_{n=1}^4 p_{n}|\varphi_{n}\rangle\langle\varphi_{n}|$ with $n\in\{1, 2, 3, 4\}$. The eigenvalues $p_{n}$ correspond to the probability that the state $\rho_{AB}$ is in the pure state $|\varphi_{n}\rangle$. The state's normalized eigenvectors $|\varphi_{n}\rangle$ can constitute an arbitrary unitary operation $E_1=\{|\varphi_{1}\rangle,  |\varphi_{2}\rangle, |\varphi_{3}\rangle, |\varphi_{4}\rangle\}$. Thus, we can attain an arbitrary two-qubit state by using an arbitrary set of probabilities and an arbitrary unitary operation.
{{The random number function $f(0,1)$ is independent
random numbers generated uniformly in a closed interval $[0,1]$. Initially, one can generate four random probabilities $p_N$ by an effective method in Ref. \cite{K.yczkowski}
\begin{align}
\begin{array}{l}
{p_1} = 1 - f(0,1) _1^{1/(N - 1)},\\
{p_2} = \left[ {1 - f(0,1) _2^{1/(N - 2)}} \right]\left( {1 - {p_1}} \right),\\
{p_k} = \left[ {1 - f\left( {0,1} \right)_k^{1/(N - k)}} \right]\left( {1 - \sum\limits_{i = 1}^{k - 1} {{p_i}} } \right),\\
\ \ \ \ \ \ \ \ \ \ \ \ \ \ \ \ \ \ \ \ \ \ \ \ \ \ \ \ \vdots \\
{p_{N - 1}} = \left[ {1 - f{{\left( {0,1} \right)}_{N - 1}}} \right]\left( {1 - \sum\limits_{i = 1}^{N - 2} {{p_i}} } \right),
\end{array}
\label{Eq.20}
\end{align}
and the last component $p_N$ is   determined as
\begin{align}
{p_N} = 1 - \sum\limits_{i = 1}^{N - 1} {{p_i}} ,
\label{Eq.21}
\end{align}
where $N$ is the dimension of the systemic state.
}}

Through the random unitary operation, we can randomly generate one 4-order real matrix ${\cal R}$ by using the random function $f(-1,1)$ with a closed interval $[-1,1]$. By  utilizing the real matrix ${\cal R}$, a random Hermitian matrix can be obtained as
\begin{align}
{\cal M}_{1}={\cal D}_{1}+({\cal U}_{1}^T+{\cal U}_{1})+i({\cal L}_{1}^T-{\cal L}_{1}),
\label{Eq.25}
\end{align}
where ${\cal D}_{1}$, ${\cal L}_{1}$ and ${\cal U}_{1}$ denote the diagonal, strictly lower- and strictly upper-triangular parts of the real matrix ${\cal R}$, respectively, and ${\cal L}_{1}^T$ depicts the transposition of the matrix ${\cal L}_{1}$.
From the calculation, we can get four normalized eigenvectors $|\varphi_{n}\rangle$ of the Hermitian matrix ${\cal{M}}_{1}$ and a random unitary operation $E_1$. Finally, we can construct the random two-qubit states.

We take {$10^6$} random states and $\hat{\sigma}_{x}, \hat{\sigma}_{y}, \hat{\sigma}_{z}$ for three MUB measurements to illustrate our results. As depicted   in Fig. \ref{figure3}(a), $U\geq U_{\rm SCB}$ holds.
Additionally, we compare SCB and Liu {\it et al.}'s bound in Fig. \ref{figure3}(b), and it is clearly shown our SCB is tighter than Liu {\it et al.}'s bound for the random states. Fig. \ref{figure3}(c) indicates that   Eq. (\ref{Eq.15}) is satisfied. Further,  it has been verified that our  OSCB is  tighter than SCB, as plotted in Fig. \ref{figure3}(d).

\subsection{Random two-qutrit states}
Next, to consider a high-dimensional scenario, we choose another group of MUB in three-dimension as
\begin{align}
\hat{{\alpha}}&=\left\{(\frac{1}{\sqrt{3}}, \frac{1}{\sqrt{3}}, \frac{1}{\sqrt{3}}), (\frac{1}{\sqrt{3}},\frac{\varepsilon^\ast}{\sqrt{3}}, \frac{\varepsilon}{\sqrt{3}}), (\frac{1}{\sqrt{3}},\frac{\varepsilon}{\sqrt{3}}, \frac{\varepsilon^\ast}{\sqrt{3}})\right\}, \nonumber\\
\hat{\beta}&=\left\{(\frac{1}{\sqrt{3}}, \frac{1}{\sqrt{3}}, \frac{\varepsilon^\ast}{\sqrt{3}}), (\frac{1}{\sqrt{3}},\frac{\varepsilon}{\sqrt{3}}, \frac{\varepsilon}{\sqrt{3}}), (\frac{1}{\sqrt{3}},\frac{\varepsilon^\ast}{\sqrt{3}}, \frac{1}{\sqrt{3}})\right\},  \nonumber\\
\hat{\gamma}&=\left\{(\frac{1}{\sqrt{3}},\frac{1}{\sqrt{3}}, \frac{\varepsilon}{\sqrt{3}}), (\frac{1}{\sqrt{3}},\frac{\varepsilon^\ast}{\sqrt{3}}, \frac{\varepsilon^\ast}{\sqrt{3}}),
(\frac{1}{\sqrt{3}},\frac{\varepsilon}{\sqrt{3}}, \frac{1}{\sqrt{3}})\right\},
\label{Eq.24}
\end{align}
where $\varepsilon={e^{2\pi i}}/{3}$ and $\varepsilon^\ast$ represents the complex conjugation of $\varepsilon$. Then, we use the same approach to construct a random two-qutrit state $\rho_{AB}$. Different from the random two-qubit state, there are 9 eigenvalues and eigenvectors to express $\rho_{AB}$. Then, we first generate a random 9-order real matrix ${\cal{K}}$ and subsequently express a random Hermitian matrix
\begin{align}
{\cal M}_{2}={\cal D}_{2}+({\cal U}_{2}^T+{\cal U}_{2})+i({\cal L}_{2}^T-{\cal L}_{2}),
\label{Eq.25}
\end{align}
where ${\cal D}_{2}$, ${\cal L}_{2}$ and ${\cal U}_{2}$ also denote the diagonal, strictly lower- and strictly upper-triangular parts of the real matrix ${\cal K}$, respectively, and ${\cal L}_{2}^T$ depicts the transposition of the matrix ${\cal L}_{2}$. One can obtain nine normalized eigenvectors of the Hermitian matrix ${\cal M}_{2}$ and a random unitary operation $E_2$. Following the similar procedure as the  generation of two-qubit random states mentioned before, the random two-qutrit states can be generated.

\begin{figure*}
\begin{minipage}{8.5cm}
\centering
\includegraphics[width=7.5cm]{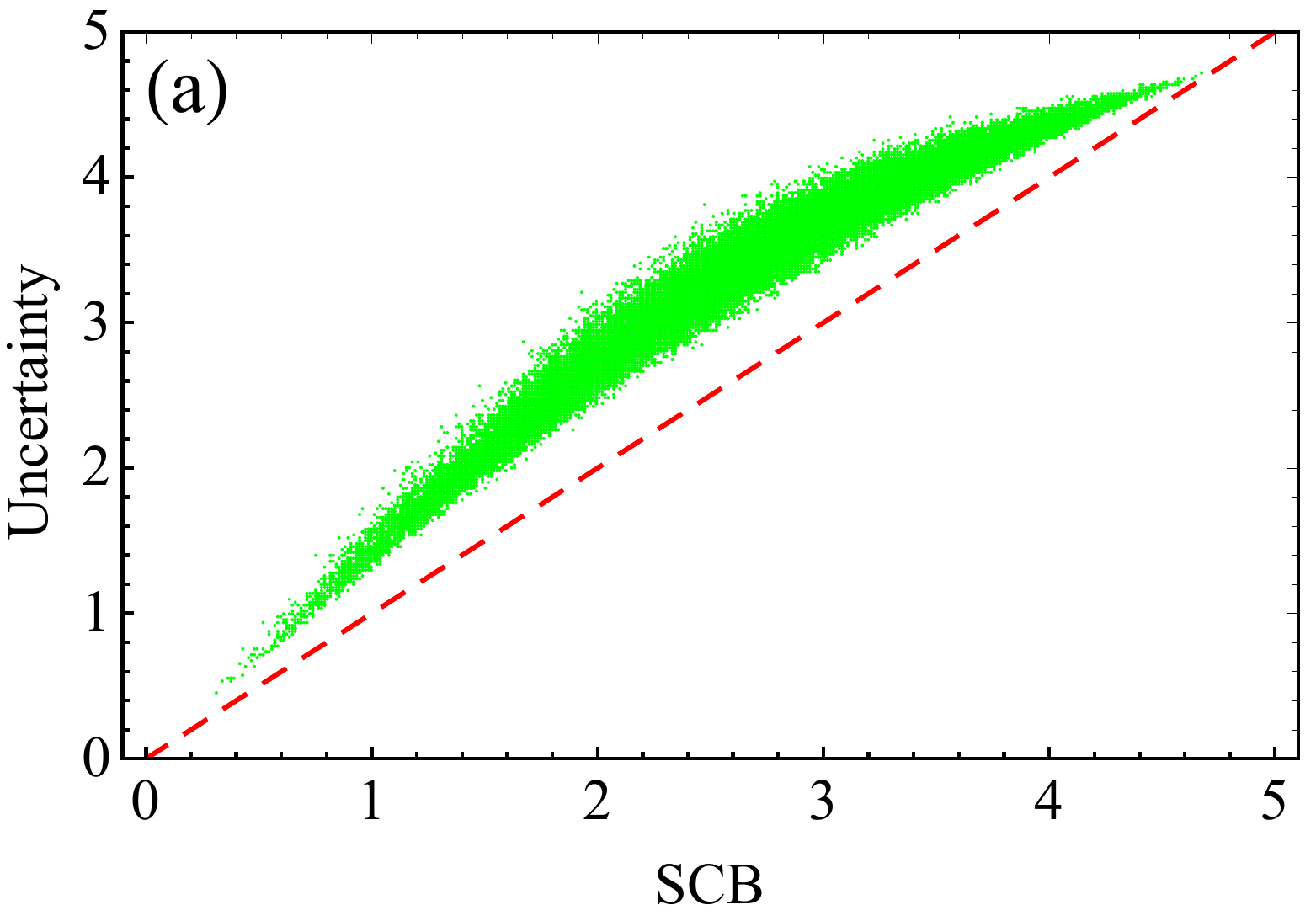}
\end{minipage}
\begin{minipage}{8.5cm}
\centering
\includegraphics[width=7.5cm]{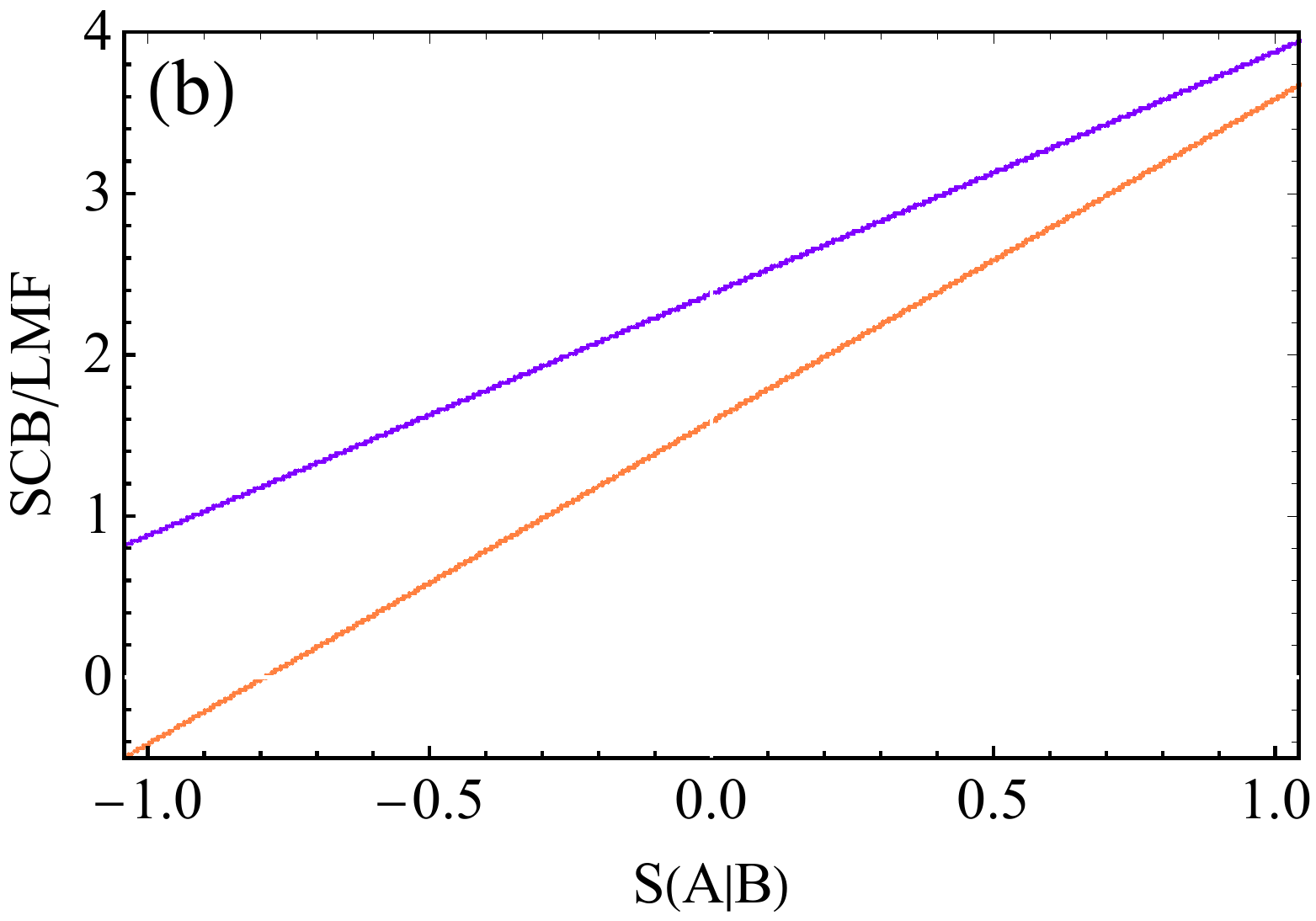}
\end{minipage}
\begin{minipage}{8.5cm}
\centering
\includegraphics[width=7.5cm]{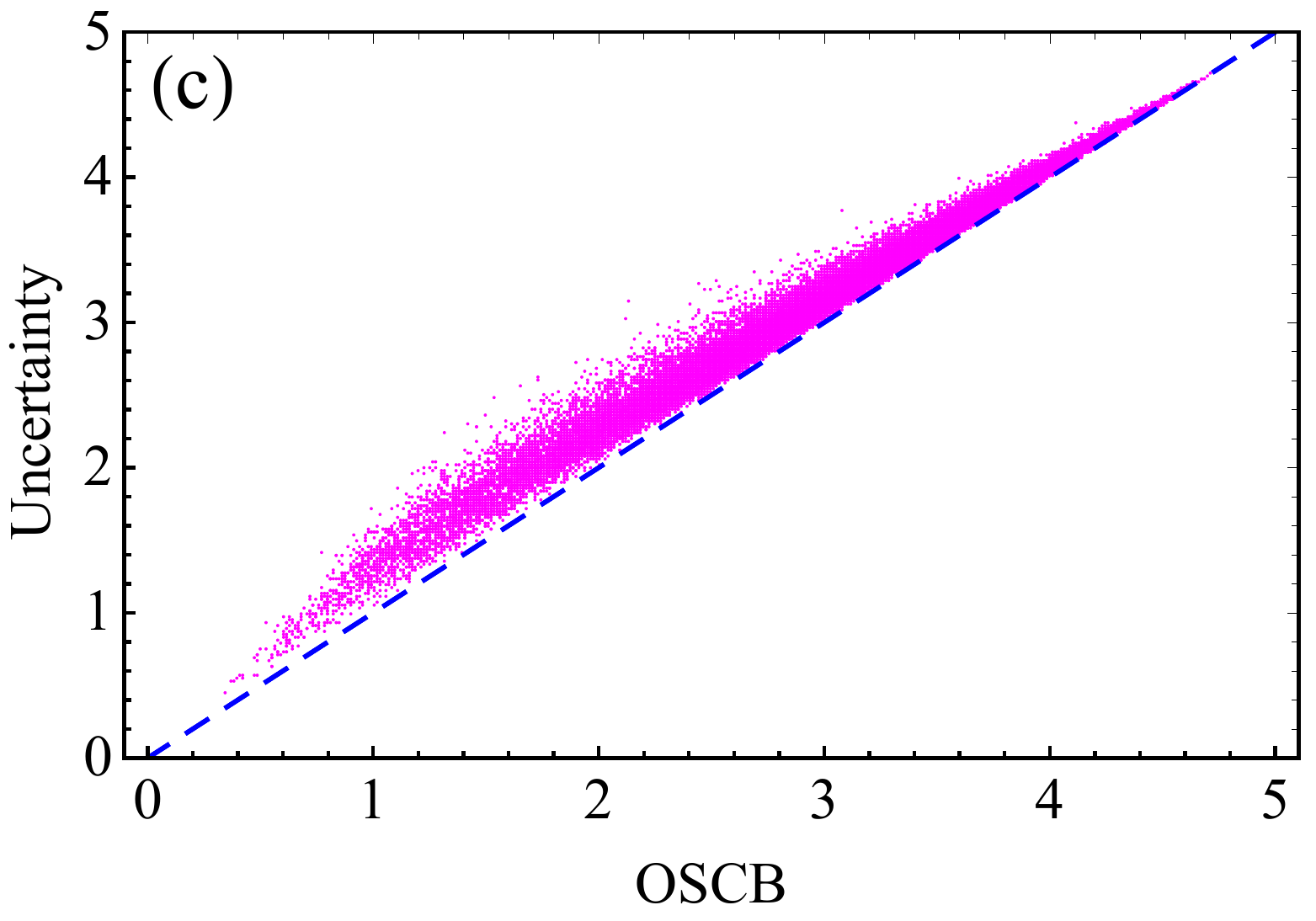}
\end{minipage}
\begin{minipage}{8.5cm}
\centering
\includegraphics[width=7.5cm]{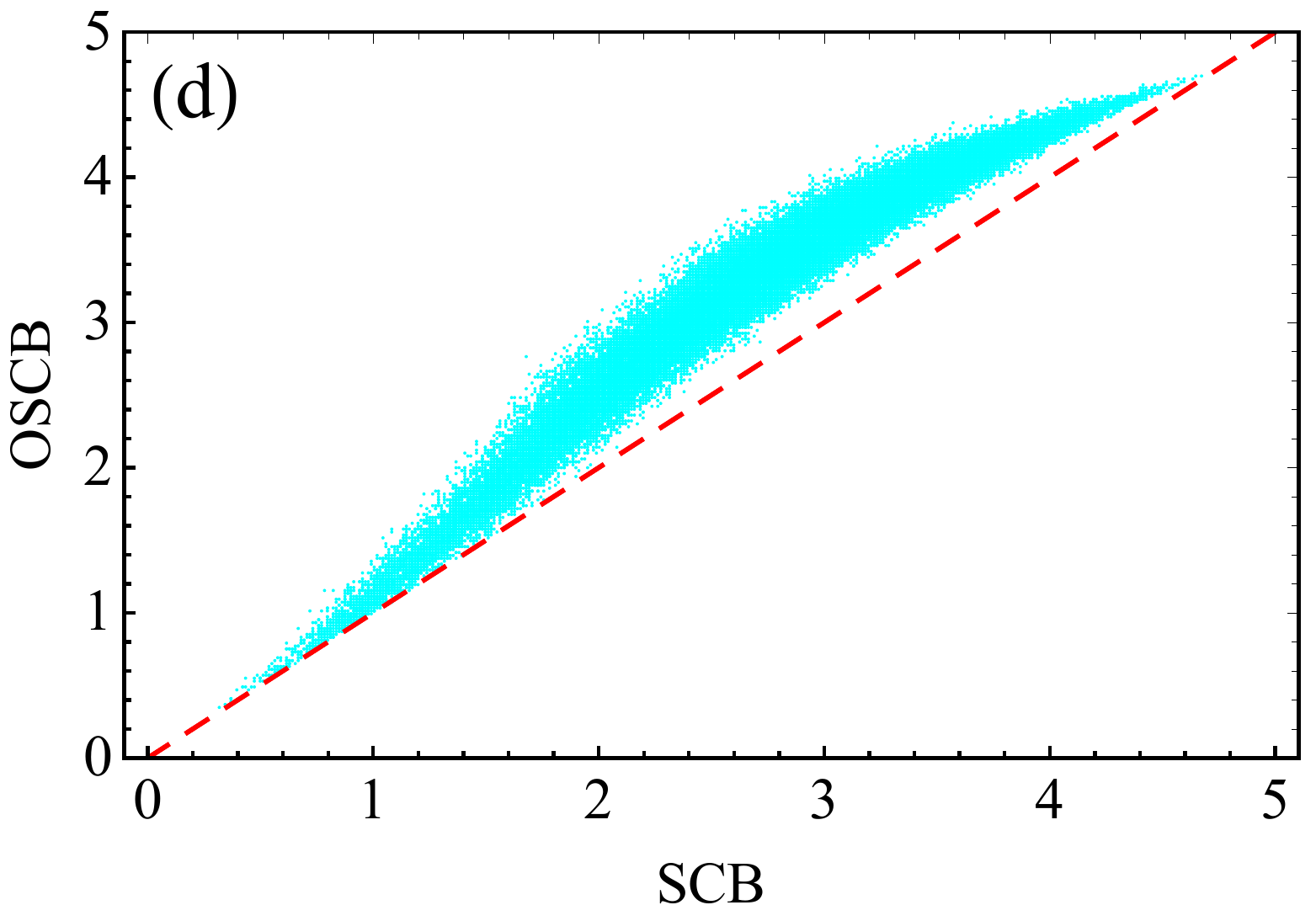}
\end{minipage}
\caption{{{The entropic uncertainty and three lower bounds under MUB multiple measurements (three three-dimensional  operators $\hat{\alpha}, \hat{\beta}, \hat{\gamma}$) for  $10^6$  randomly generated two-qubit states. Graph (a): $x$ axis represents our simply constructed bound (SCB) in Eq. (\ref{Eq.6}) and
$y$ axis denotes the entropic uncertainty in Eq. (\ref{Eq.6}). The red dashed line is the proportional function with a slope of unity, displaying that the
uncertainty equals to SCB. Graph (b): the orange points stand  for Liu {\it et al.}'s bound (LMF) in Eq. (\ref{Eq.5}) and the purple points denote our proposed our proposed bound (SCB),  and $S(A|B)$ is the conditional von Neumann entropy of state $\rho_{AB}$.   Graph (c):   $x$ axis represents our
optimized simply constructed bound (OSCB) in Eq. (\ref{Eq.12}) and $y$ axis represents the entropic uncertainty in Eq. (\ref{Eq.6}), and the blue dashed line is the
proportional function with a slope of unity, displaying that the uncertainty equals to OSCB.  Graph (d): SCB represents the lower bound  in Eq. (\ref{Eq.6}) and OSCB denotes the lower bound in Eq. (\ref{Eq.12}). The red dashed line is the proportional function with a slope of unity, displaying that OSCB equals to SCB.}}}
\label{figure4}
\end{figure*}

To compare our results with the ones reported in Ref. \cite{S. Liu}, we take {$10^6$} random three-qutrit states and three groups of three-dimension MUB measurements ($\hat{\alpha},\hat{\beta},\hat{\gamma}$), as shown in  Fig. \ref{figure4}. From the figure, we
conclude that: (i) the uncertainty is not less than our proposed SCB and OSCB, i.e., $U\geq U_{\rm SCB}$ and $U\geq U_{\rm OSCB}$ as shown in Figs. \ref{figure4}(a) and \ref{figure4}(c), verifying Eqs. (\ref{Eq.6}) and (\ref{Eq.12}); (ii) Our SCB is tighter than Liu {\it et al.}'s bound and our presented OSCB is stronger than both the SCB and LMF bound, i.e., $U_{\rm OSCB} \geq U_{\rm SCB}\geq U_{\rm LMF}$ is satisfied, as shown in Figs. \ref{figure4}(b) and  \ref{figure4}(d).
%
%

\section{Conclusion}
To conclude, we derived the simply constructed bound and the optimized simply constructed bound of entropic uncertainty relations in the multiple measurements architecture. We proved that our SCB and OSCB outperform the  one proposed by Liu {\it et al.} for arbitrary MUB multiple measurements. We considered Werner-type states, Bell-diagonal states, random two-qubit states and random two-qutrit states as illustrations to verify and support our results.
By considering the mutual information for OSCB, we built the intrinsic relationship the uncertainty relation and systemic correlation; this reflects that the nature of uncertainty may be determined by the systemic correlation, containing both the classical and quantum components. More specifically, the total correlation might determine the dynamic features of uncertainty. Notably, multiple measurement relations with quantum memory may be treated as the criteria for entanglement in a complementary way {\cite{S. Liu}}. Besides, the experimental realization of two-measurement settings has been  demonstrated in all-optics platforms {\cite{C. F. Li,R. Prevedel,ding}}, and the multiple measurement uncertainty relation for a bipartite system has also been implemented in a nuclear spin system in Ref. {\cite{H. Y. Wang}}. Therefore, the
 current proposal can be expected to be experimentally implemented in the near future.

Hence, we believe that our tighter entropic uncertainty relations would be beneficial in shedding light on quantum precision measurements and be of fundamental importance in measurement-based quantum key distribution protocols.

\section{Acknowledgements}
This work was supported by the National Science Foundation of China under Grant Nos. 12075001, 61601002, 11875167 and 11575001, Anhui Provincial Natural Science Foundation (Grant No. 1508085QF139) and  CAS Key Laboratory of Quantum Information (Grant No. KQI201701).

B.F.X. and F.M. contributed equally to this work.

\newcommand{\RMP}{\emph{Rev. Mod. Phys.} }
\newcommand{\PRA}{{Phys. Rev. A} }
\newcommand{\PRB}{\emph{Phys. Rev. B} }
\newcommand{\PRE}{\emph{Phys. Rev. E} }
\newcommand{\PRD}{\emph{Phys. Rev. D} }
\newcommand{\APL}{\emph{Appl. Phys. Lett.} }
\newcommand{\NJP}{\emph{New J. Phys.} }
\newcommand{\JPA}{\emph{J. Phys. A} }
\newcommand{\JPB}{\emph{J. Phys. B} }
\newcommand{\OC}{\emph{Opt. Commun.} }
\newcommand{\PLA}{\emph{Phys. Lett. A} }
\newcommand{\EPJD}{\emph{Eur. Phys. J. D} }
\newcommand{\NP}{\emph{Nat. Phys.} }
\newcommand{\NC}{\emph{Nat. Commun.} }
\newcommand{\EPL}{\emph{Europhys. Lett.} }
\newcommand{\AoP}{\emph{Ann. Phys.} }
\newcommand{\ADP}{\emph{Ann. Phys. (Berlin)} }
\newcommand{\QIC}{\emph{Quantum Inf. Comput.} }
\newcommand{\QIP}{\emph{Quantum Inf. Process.} }
\newcommand{\CPB}{\emph{Chin. Phys. B} }
\newcommand{\IJTP}{\emph{Int. J. Theor. Phys.} }
\newcommand{\IJMPB}{\emph{Int. J. Mod. Phys. B} }
\newcommand{\PR}{\emph{Phys. Rep.} }
\newcommand{\SR}{\emph{Sci. Rep.} }
\newcommand{\LPL}{\emph{Laser Phys. Lett.} }
\newcommand{\OEE}{\emph{Opt. Exp.} }
\newcommand{\IJQI}{\emph{Int. J. Quantum Inf.} }

\end{document}